\documentstyle[aps,prl,preprint,floats,epsfig]{revtex}

\begin{document}

\preprint{\tighten\vbox{\hbox{\hfil CLNS 99/1626}
                        \hbox{\hfil CLEO 99-7}
}}

\title{Charged Track Multiplicity in $B$ Meson Decay}

\author{CLEO Collaboration}
\date{\today}

\maketitle
\tighten

\begin{abstract} 
We have used the CLEO II detector to study the multiplicity of 
charged particles in the decays of $B$ mesons produced at the $\Upsilon(4S)$
resonance.  Using a
sample of $1.5{\times}10^6$ $B$ meson pairs, we find the mean inclusive 
charged particle multiplicity to be $10.71\pm{0.02}_{-0.15}^{+0.21}$
for the decay of the pair.  
This corresponds to a mean multiplicity of 
$5.36\pm{0.01}_{-0.08}^{+0.11}$ for a single $B$ meson.
Using the same data sample, we have
also extracted the mean multiplicities in semileptonic and nonleptonic
decays.  We measure a mean of $7.82\pm{0.05}_{-0.19}^{+0.21}$ charged
particles per $B\bar{B}$ decay when both mesons decay semileptonically.
When neither $B$ meson decays semileptonically, we measure a mean charged
particle multiplicity of $11.62\pm{0.04}_{-0.18}^{+0.24}$ per 
$B\bar{B}$ pair.
\end{abstract}
\newpage

{
\renewcommand{\thefootnote}{\fnsymbol{footnote}}

\begin{center}
G.~Brandenburg,$^{1}$ A.~Ershov,$^{1}$ Y.~S.~Gao,$^{1}$
D.~Y.-J.~Kim,$^{1}$ R.~Wilson,$^{1}$
T.~E.~Browder,$^{2}$ Y.~Li,$^{2}$ J.~L.~Rodriguez,$^{2}$
H.~Yamamoto,$^{2}$
T.~Bergfeld,$^{3}$
% ...
\end{center}
 
\small
\begin{center}
$^{1}${Harvard University, Cambridge, Massachusetts 02138}\\
$^{2}${University of Hawaii at Manoa, Honolulu, Hawaii 96822}\\
$^{3}${University of Illinois, Urbana-Champaign, Illinois 61801}\\
% ...
\end{center}
\setcounter{footnote}{0}
}
\newpage

\section{INTRODUCTION}

Measurements of the charged track mutliplicity distribution in $B$ meson
decay are used to constrain unmeasured or poorly measured branching 
fractions in Monte Carlo simulations so that generated event samples more 
closely represent actual data.  The CLEO Monte Carlo parameterization of
$B$ meson decays has been tuned to agree with our measurements and our
model is used by other experimental groups~\cite{CDF}.
Charged particle multiplicity in heavy meson decay has been studied by several
groups~\cite{CLEOI}\cite{MARKIII}\cite{ACCMOR}.  
In this paper we present a measurement of the charged particle 
multiplicity in inclusive $B\bar{B}$ decays that is an improvement
over our previous result~\cite{CLEOI}.  
We also present improved measurements of
the charged particle multiplicities in semileptonic and nonleptonic decays. 

For clarity, we use the term ``observed multiplicity'' to denote the
number of well reconstructed charged particle tracks in a given event.
We use the term ``decay multiplicity'' to denote the number of
$e^{\pm}$, $\mu^{\pm}$, $\pi^{\pm}$, $K^{\pm}$ and $p^{\pm}$ 
that come from the decay of the 
primary $B$ mesons and also from the subsequent decays of any secondary or
tertiary particles other than neutrons, $K_L$, or $\pi^0$.  The decay
multiplicity excludes any tracks produced through interactions with the
detector or surrounding material.
Not all charged decay products will result in reconstructed tracks, 
and not all observed tracks come from the primary decay, so the 
observed multiplicity may be less than, equal to, or greater than
the decay multiplicity for a given event.

\section{INCLUSIVE MULTIPLICITY MEASUREMENT}
The CLEO detector is located at
the Cornell Electron Storage Ring, a high luminosity $e^+e^-$ collider 
operated at or near the $\Upsilon(4S)$ resonance.
The results presented here are derived from a sample of 1.4~fb$^{-1}$, 
corresponding to $1.5{\times}10^6$ $B$ meson pairs, collected 
with the CLEO II detector~\cite{CD}.  
Charged particle tracks are measured by
cylindrical wire drift chambers inside a 1.5~T superconducting solenoid.
A CsI crystal calorimeter is also inside the magnet, and 
energy deposition information from both the calorimeter and the drift chamber 
is used for particle
identification.  Muon counters are layered in the steel yoke surrounding 
the coil.

To obtain a clean sample of candidate $B\bar{B}$ events, we select
hadronic events by requiring that an event have three or more 
reconstructed tracks, energy deposition in the calorimeter greater than
15\% of the center of mass energy and an event vertex consistent with 
the interaction region.  For additional background suppression, 
the total reconstructed event energy,  including charged and neutral 
particles, is required to be between 4~GeV 
and 12~GeV, and the total reconstructed vector 
momentum of the event is required 
to have a magnitude less than 3~GeV/$c$.
This hadronic event sample contains events from 
both $B\bar{B}$ and continuum processes 
such as $q\bar{q}$ and $\tau^+\tau^-$ production.
We remove the continuum contribution by rescaling and
subtracting the observed multiplicity distribution of 
a separate 0.7~fb$^{-1}$ data sample collected 65~MeV/$c^2$ 
below the $\Upsilon(4S)$ resonance.

To be counted in our observed multiplicity, 
drift chamber tracks are required to be well reconstructed and consistent 
with having originated from 
the event vertex.  Tracks must not be within 25.8~degrees of the 
$e^+e^-$ beam axis.  
Once the event selection, continuum subtraction, and track
selection are completed, we count the selected tracks in each 
event to obtain the {\it observed} charged track multiplicity distribution 
in Fig.~\ref{fig1}.  There are events with fewer than three selected
tracks because not all reconstructed tracks pass the track selection
criteria.

\vspace{0.1in}
\makebox[368bp]{\epsfysize=3.5in \epsfxsize=3.5in  \epsffile{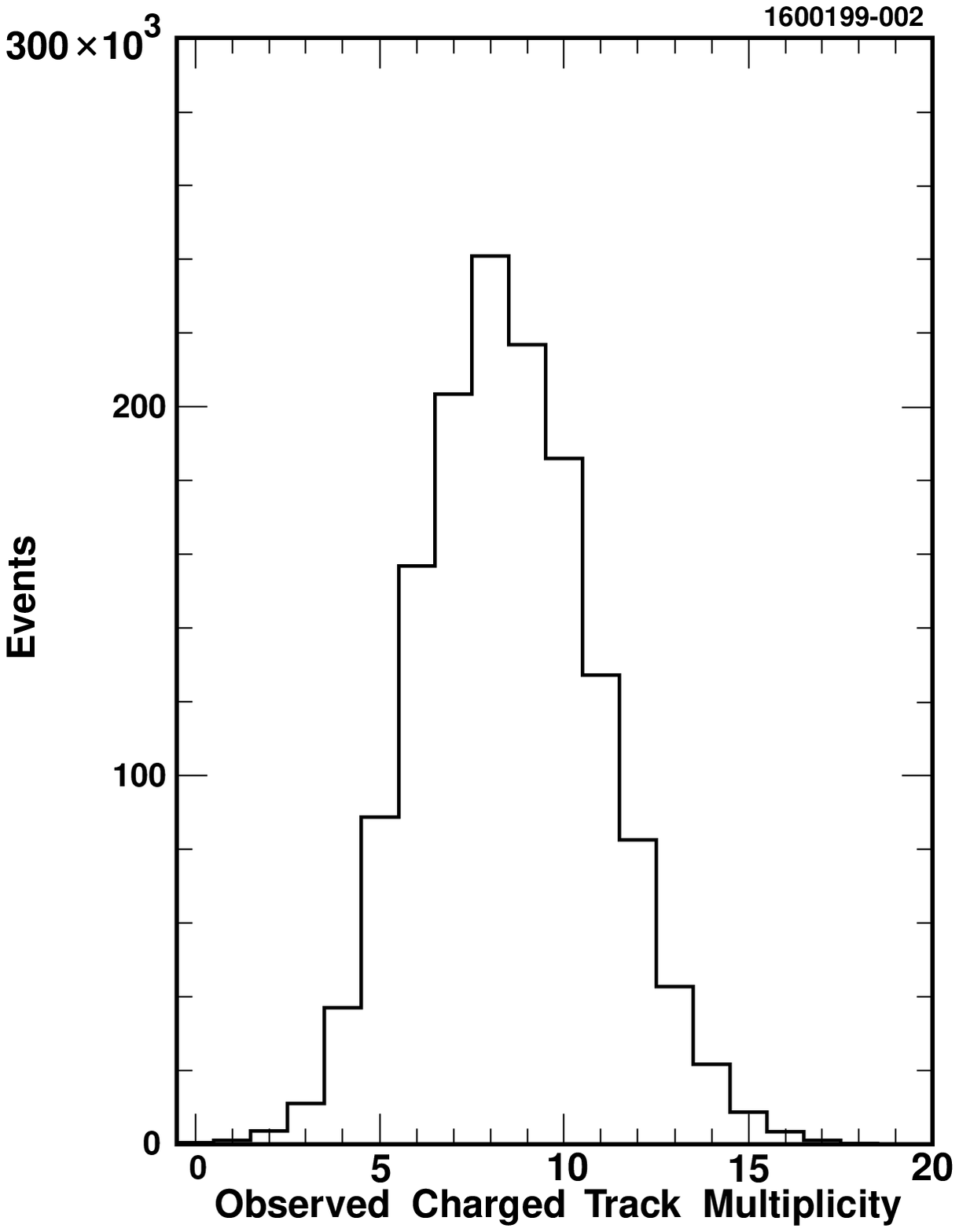} }
\begin{figure}[ht]
\vspace{0.1in}
\caption{Observed charged track multiplicity for $B\bar{B}$ events.}
\label{fig1}
\end{figure}

To obtain the true decay multiplicity distribution 
from the observed multiplicity distribution, we must
account for detector effects.
If the number of events with observed multiplicity $j$ is $O_j$ and 
the number of events with decay multiplicity $i$ is $D_i$, these
quantities are related by Eqn.~\ref{E:migration}.

\begin{equation}
O_j = \sum_{i=2,4,...}^{n}\epsilon_{ij} D_i
\label{E:migration} 
\end{equation} 

\noindent 
where $\epsilon_{ij}$ 
is the probability that an event with decay multiplicity $i$ will 
be reconstructed with observed multiplicity $j$.  We have assumed
that charge is conserved, so the index $i$ can only take even
values.
In principle, there
can be events with decay multiplicity of zero, where two neutral $B$ 
mesons decay to all neutral final states.  However, we do not include
zero decay multiplicity events in our analysis both because of the very
low branching ratio for such events 
and also because our event selection criteria make
detection of such events extremely unlikely.
The upper bound in Eqn.~\ref{E:migration} is, in principle, the maximum
decay multiplicity in a $B\bar{B}$ event, which is not known.  We vary
the maximum decay multiplicity in our analysis as described below.
The fact that the sum of $D_i$ 
and the sum of $O_j$ are both equal to the total number 
of events is used to constrain
the values of the $D_i$, as expressed in 
Eqn.~\ref{E:constraint}.

\begin{equation}
\sum_{i=2,4,...}^{n} D_i = \sum_{j=0}^{m} O_j
\label{E:constraint} 
\end{equation} 

\noindent 
where the upper bound $m$ is the maximum value of our observed
multipicity which is 20.

The coefficients $\epsilon_{ij}$ in Eqn.~\ref{E:migration}
are obtained from Monte Carlo
simulation and depend primarily on the detector's track finding efficiency
and also on the probability of producing 
extra charged particles that pass the track selection cuts and that
are counted.  While these coefficients depend on accurate 
simulation of detector response and processes such as photon
conversion and decays in flight, they do not depend significantly
on the exact tuning of the branching fractions or the decay 
multiplicity distribution in the simulation.

\begin{figure}[ht]
\vspace{0.1in}
\makebox[368bp]{\epsfysize=3.5in \epsfxsize=3.5in \epsffile{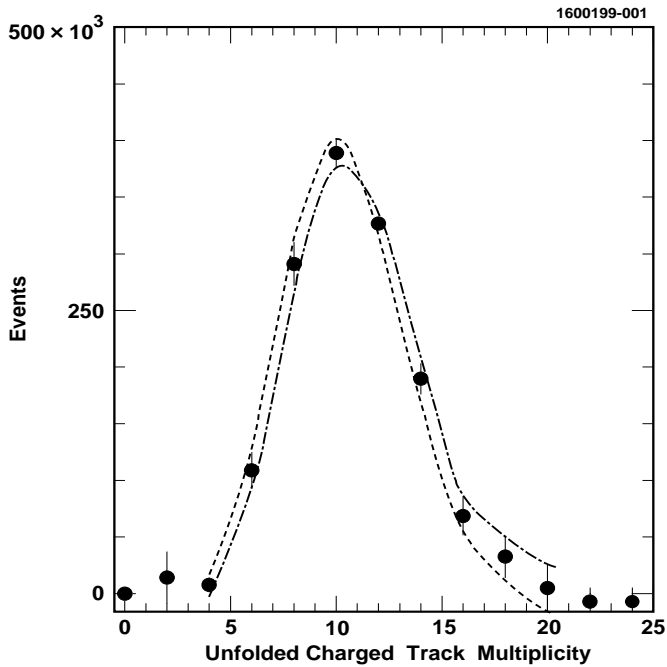} }
\vspace{0.1in}
\caption{Unfolded charged track decay multiplicity for $B\bar{B}$ events.
The dashed lines represent the high multiplicity and low multiplicity
fluctuations in the fit.}
\label{fig2}
\end{figure}

The parameters $D_i$ in Eqn.~\ref{E:migration} are determined 
by a $\chi^2$ fit.  
This fit unfolds the detector effects to give
the decay 
multiplicity distribution of events that pass the event selection
criteria.  These selection 
criteria are biased against very low decay multiplicity
events, particularly because of the requirement of three or more 
reconstructed tracks.  We remove the event selection bias using 
Monte Carlo simulation to determine the probability for events of a given
decay multiplicity to pass the event selection cuts.  
After unfolding detector and reconstruction effects,
we obtain the decay multiplicity distribution in 
Fig.~\ref{fig2}.  The error bars represent the statistical 
uncertainty in both the $\chi^2$ fit and the event bias correction,
but do not include systematic errors.  The dashed lines represent
the high multiplicity and low multiplicity statistical fluctuations
in the fit.
The large error bar on the $i~=~2$ point is due to the event bias
correction.

From the distribution in Fig.~\ref{fig2}, 
we obtain a mean of $10.71\pm{0.02}$ charged
particles for inclusive $B\bar{B}$ decay, where the error is statistical
only.  The points with $i~>~20$ show no evidence for events with such
high multiplicities.  We have tested our fitting procedure by varying
the maximum decay multiplicity included in the fit between 20 and 28.  
The fit is stable
and the unfolded mean decay multiplicity does not change significantly when
decay multiplicities of 22 and higher are included or excluded from the fit.

The most important systematic effect in this analysis is the
accuracy of
modeling the detector's track finding efficiency.  Our studies
indicate that the overall efficiency for finding a single track is known
to within $\pm{1\%}$ for tracks with momentum greater than $250$~MeV/$c$ and
with decreasing accuracy as track momentum decreases.  
The uncertainty in single track finding efficiency gives a $_{-0.9}^{+1.6}\%$ 
uncertainty in the measured mean decay multiplicity.
Removal of the event selection bias shifts the measured 
mean decay multiplicity by $+0.7\%$, which we also take as the systematic 
uncertainty for event selection bias.  
Because the track finding efficiency depends on the momentum of the
track, we account for the uncertainty due to this dependence.
This analysis uses all tracks without regard to momentum.  When
we add a track
selection cut requiring a reconstructed momentum of at least $150$~MeV/$c$,
we observe a shift in the decay multiplicity of $-0.6\%$. 
We assign an additional $\pm{0.6\%}$ uncertainty due to this momentum 
dependence.  
Charged pions produced by the decays of $K_S$ will have 
lower track finding efficiency than average charged pions, so our
result depends on the rate of $K_S$ production in our simulation.
Based on our studies of inclusive $K_S$ production in $B$ meson 
decay, we assign
a systematic uncertainty of $^{+0.5}_{-0.3}\%$ from this source.
Interactions of neutral and
charged decay products with the detector material produce additional
charged tracks, some of which satisfy the track selection criteria.
We study the effect of these extra tracks by varying the rates of
photon conversion and hadronic interactions in our Monte Carlo sample.
Misidentification of these extra tracks contributes an additional 
$\pm{0.3\%}$ systematic uncertainty.
Extra tracks also come from the decay of particles in 
the detector volume, adding another $\pm{0.1\%}$ uncertainty.  
Additional uncertainty comes from contamination by non-$B\bar{B}$ events, 
most notably from beam-gas interactions.  
We study the effect of 
non-$B\bar{B}$ events by varying the size of the continuum subtraction
by $1\%$, which is the uncertainty in our measured luminosity,
and we find a $\pm{0.2\%}$ uncertainty in our mean decay multiplicity.  

We have looked
for other potential systematic effects by varying our track selection 
cuts but do not observe any significant change in mean decay multiplicity.
When all systematic errors are added in quadrature, we obtain a total
systematic uncertainty of $_{-1.4}^{+1.9}\%$, which gives a final
result of $10.71\pm{0.02}_{-0.15}^{+0.21}$
for the mean inclusive charged particle decay multiplicity in the 
decay of a $B\bar{B}$ pair.
Systematic errors are summarized in Table~\ref{tab1}.

The results presented here are a significant improvement over the
previous CLEO result~\cite{CLEOI}.  The earlier analysis found a
mean multiplicity of $11.5\pm{0.2}\pm{0.4}$.  In addition to using
a much larger sample of $B\bar{B}$ events, the present analysis 
uses a substantially different method.  The former analysis treated
the coefficients $\epsilon_{ij}$ in Eqn.~\ref{E:migration} as a 
matrix and used matrix inversion to solve for the $D_i$.  Such 
matrix inversion is very sensitive to singularities and can produce
unstable results.  In the current analysis, we do not attempt to 
invert the $\epsilon_{ij}$ matrix, but instead we perform a more
stable $\chi^2$ fit for the parameters $D_i$.

\begin{center}
\begin{table}[htb]
\begin{tabular}{|l|c|}\hline
Systematic Error Source   & Effect on Mean Decay Multiplicity    \\ 
			  & (Charged Tracks)                     \\ \hline
Overall Track Finding Efficiency            & $_{-0.09}^{+0.17}$   \\ \hline
Event Selection Bias                        & $\pm{0.07}$          \\ \hline
Low Momentum Tracking Efficiency            & $\pm{0.06}$          \\ \hline
$K_S$ Modeling                              & $_{-0.03}^{+0.05}$   \\ \hline
Hadronic Interactions and Photon Conversion & $\pm{0.03}$          \\ \hline
Contamination by Non $B\bar{B}$ Events      & $\pm{0.02}$          \\ \hline
Decays in Flight                            & $\pm{0.01}$          \\ \hline
All Other Sources                           & $\pm{0.01}$        \\ \hline
Total Systematic Error                      & $_{-0.15}^{+0.21}$   \\ \hline
\end{tabular}
\caption{Systematic Errors for the Inclusive Decay Multiplicity Measurement
for $B\bar{B}$ Pairs.}
\label{tab1}
\end{table}
\end{center}

\section{SEMILEPTONIC AND NONLEPTONIC MULTIPLICITY MEASUREMENT}

We have further analyzed the same data sample to measure multiplicities 
separately for semileptonic and nonleptonic decays.  We define 
semileptonic to include only decays of $B$ mesons into an electron or muon
plus a neutrino and any number of hadrons.  All other decays are classified as 
nonleptonic.  Decays involving tau leptons 
are counted as nonleptonic because approximately 65\% of 
taus decay hadronically~\cite{TAUBR}
and such events cannot be distinguished easily from purely hadronic $B$ decays.

We use the same event and track selection criteria as in the inclusive
analysis and sort the events by the number of leptons identified.
We identify electrons by combining information on energy deposition in 
the drift chamber, the shape of the shower observed in the calorimeter,
and the ratio of the calorimeter energy to the track momentum.
Muons are required to traverse at least three pion nuclear
interaction lengths of material.  We also require all lepton candidates 
to have momentum between $1.4$~GeV/$c$ and $2.5$~GeV/$c$ to suppress false
lepton identification and secondary leptons.

Once we have sorted the events by the number of leptons found, we
correct for misidentified and secondary leptons.
We move these misidentified events from the sample with one identified 
lepton to the sample of events with no identified lepton
by using the observed multiplicity distributions of Monte Carlo 
generated events that are similarly misidentified in the reconstruction.
The fake and secondary observed 
multiplicity distributions are scaled using our best 
estimates of the fake and secondary lepton rates and are added to the
sample with no identified lepton and subtracted from the sample with one
identified lepton.
Uncertainties in the rates and observed multiplicity distributions of
these events are treated as systematic errors.

We use the same methods as in the inclusive analysis to extract the
mean decay multiplicity from the corrected observed multiplicity
distributions for the samples with zero or one identified lepton.  
For events with no reconstructed leptons, we obtain a mean decay 
multiplicity of
$11.02\pm{0.01}$, and for the sample with one reconstructed lepton, we
find a mean of $9.28\pm{0.02}$ charged tracks, where the errors are
statistical only.

Not all semileptonic decays will produce a detected lepton because not all
electrons and muons will enter the fiducial tracking volume of the detector
and pass the track selection and lepton identification 
criteria.  
To obtain the true decay multiplicities for 
semileptonic and nonleptonic $B\bar{B}$ decays, we use simulation to unfold 
the migration
between the number of leptons generated by the decay of the $B$ mesons and
the number of leptons actually identified.
If the migration probability for a given event were independent 
of the event's decay multiplicity, we could write:

\begin{equation}
M^i = a^i M_{n-n}  +  b^i M_{s-s},
\label{E:lepmigration}
\end{equation} 

\noindent where the $M^{i}$ are the mean decay 
multiplicities of the sample with $i$ 
identified leptons, $M_{n-n}$ and $M_{s-s}$ are the mean decay 
multiplicities of events where both $B$ mesons decay nonleptonically
and semileptonically, respectively, 
and $a^i$ and $b^i$ are migration coefficients
determined from Monte Carlo simulation.
These multiplicities include the lepton tracks.

There are many factors that affect the number of leptons that will be
reconstructed for a given event.  Lepton identification depends strongly 
on particle momentum not only because of the explicit momentum requirements
but also because a particle's momentum determines the probability that it
will penetrate the muon chambers sufficiently to be classified as a muon
candidate.
Because of phase space limitations, 
events with one or more high momentum (above~$1.4$~GeV/$c$)
particles will tend to have fewer charged tracks
than events without a high momentum particle, so events with an identified
lepton tend to have lower multiplicities than those where the lepton was
not identified.


In addition to assuming that migration 
probability is independent of decay multiplicity, Eqn.~\ref{E:lepmigration}
also assumes that the mean decay 
multiplicity of events where only one $B$ meson decays 
semileptonically is simply the average of $M_{n-n}$ and $M_{s-s}$.  This
assumption would be valid if the mean true decay multiplicities of $B^-$
and $B^0$ mesons were equal, but this may not be the case.
To account for the possibility of unequal
$B^-$ and $B^0$ multiplicities and the effect of multiplicity
dependent migration, we can introduce another term into 
Eqn.~\ref{E:lepmigration}:

\begin{equation}
M^i + \Delta^i= a^i M_{n-n}  +  b^i M_{s-s}.
\label{E:fixedmigration}
\end{equation} 

The correction terms $\Delta^i$ and migration coefficients are determined
from Monte Carlo simulation, 
and we can solve Eqn.~\ref{E:fixedmigration} for 
$M_{n-n}$ and $M_{s-s}$.  We obtain mean true decay multiplicities of 
$11.62\pm{0.04}$ charged particles 
for events where neither $B$ meson decays semileptonically and 
$7.82\pm{0.05}$ charged particles for events where both $B$ mesons 
decay semileptonically, where the errors are statistical and include
analytic propagation of the uncertainty in each of the parameters in 
Eqn.~\ref{E:fixedmigration}.

All of the systematic uncertainties that affect the inclusive analysis are
also present in these semileptonic and nonleptonic results.  This 
$^{+1.9}_{-1.4}\%$ uncertainty in unfolding the decay multiplicity from
the observed multiplicity is the largest systematic effect
for these results.  The next largest uncertainty comes from
lepton identification, including inefficiencies, false particle
identification, and acceptance of secondary leptons not directly produced
in the $B$ meson decay.
We have studied these effects by varying all of our lepton identification
criteria.  We have also varied the rates of fake and secondary lepton
identification directly in simulation.  All of the lepton identification
systematics combine to give a $\pm 0.06$ charged track uncertainty in the 
nonleptonic result and a corresponding uncertainty of $\pm 0.11$ 
charged tracks in the mean decay multiplicity
of semileptonic decays.  Our result is also sensitive to the 
modeling of semileptonic decays in our simulation, including the inclusive
rate of semileptonic decays and the exclusive branching fractions of 
semileptonic decays that produce the various resonances of the $D$ meson.
We have varied both the inclusive and exclusive branching fractions within
their uncertainties and find $\pm 0.05$ uncertainty in the nonleptonic
mean decay 
multiplicity and $\pm 0.08$ uncertainty for the semileptonic result.
Uncertainty in the observed 
multiplicity of events with fake and secondary leptons 
contributes $\pm 0.06$ charged tracks uncertainty to the semileptonic
result.  The systematic uncertainties 
are summarized in Table~\ref{Ta:lsystematic}.

Combining all errors, we find mean multiplicities of 
$11.62\pm{0.04}^{+0.24}_{-0.18}$ charged tracks when both $B$ mesons
decay nonleptonically and $7.82\pm{0.05}^{+0.21}_{-0.19}$ charged
tracks when both $B$ mesons decay semileptonically.
As a further check on our systematic errors, we have repeated this
analysis by using only electron identification
or only muon identification.  In both cases, the results
are statistically consistent with the means stated above and with
each other.  Our present results are a significant improvement
over the previous CLEO results~\cite{CLEOI}, in which mean multiplicities
of $12.6\pm{0.4}\pm{0.4}$ and $8.2\pm{0.7}\pm{0.4}$ were obtained for
events where both $B$ mesons decay nonleptonically and semileptonically,
respectively.

\begin{table} 
\centering\begin{tabular}{|l|c|c|} \hline
Systematic Error Source & Nonleptonic Events & Semileptonic Events \\ 
                        & (Charged Tracks)    & (Charged Tracks)     \\ \hline
Inclusive Unfolding     & $^{+0.23}_{-0.16}$  & $^{+0.15}_{-0.11}$   \\ \hline
Lepton Identification   & $\pm 0.06$          & $\pm 0.11$           \\ \hline
Branching Fractions     & $\pm 0.05$          & $\pm 0.08$           \\ \hline
Fake and Secondary Leptons & $\pm 0.01$  & $\pm 0.06$           \\ \hline
All Other Sources       & $<\pm 0.01$         & $\pm 0.01$           \\ \hline
Total Uncertainty       & $^{+0.24}_{-0.18}$  & $^{+0.21}_{-0.19}$   \\ \hline
\end{tabular}
\vspace{0.15in}
\caption[Systematic errors for the charged track multiplicity 
in semileptonic and nonleptonic events.]
{Systematic Error for the Charged Track Mutliplicity in  
Nonleptonic and Semileptonic Events.  
} \label{Ta:lsystematic}
\end{table}

\section{ACKNOWLEDGEMENTS}

We gratefully acknowledge the effort of the CESR staff in providing us with
excellent luminosity and running conditions.
J.R. Patterson and I.P.J. Shipsey thank the NYI program of the NSF, 
M. Selen thanks the PFF program of the NSF, 
M. Selen and H. Yamamoto thank the OJI program of DOE, 
J.R. Patterson, K. Honscheid, M. Selen and V. Sharma 
thank the A.P. Sloan Foundation, 
M. Selen and V. Sharma thank the Research Corporation, 
F. Blanc and S. von Dombrowski thank the Swiss National Science Foundation, 
and H. Schwarthoff and E. von Toerne thank 
the Alexander von Humboldt Stiftung for support.  
This work was supported by the National Science Foundation, the
U.S. Department of Energy, and the Natural Sciences and Engineering Research 
Council of Canada.


\begin{thebibliography}{99}

\bibitem{CDF}
See, for example, CDF Collaboration, F.~Abe {\it et al.}, 
Phys. Rev. D {\bf 50}, 2966 (1994), and
D0 Collaboration, B.~Abbot {\it et al.}, 
hep-ex/9905024 (1999).

\bibitem{CLEOI}
CLEO Collaboration, M.S.~Alam {\it et al.},
Phys. Rev. Lett. {\bf 49}, 357 (1982).

\bibitem{MARKIII}
MARK III Collaboration, D.~Coffman {\it et al.}, 
Phys. Lett. B {\bf 263}, 135 (1991).

\bibitem{ACCMOR}
ACCMOR Collaboration, S.~Barlag {\it et al.}, 
Z. Phys. C {\bf 55}, 383 (1992).

\bibitem{CD}
CLEO Collaboration, Y.~Kubota {\it et al.}, 
Nucl. Instrum. Methods Phys. Res., Sect. A {\bf 320}, 66 (1992).

\bibitem{TAUBR}
C.~Caso {\it et al.}, 
European Phys. J. {\bf C3}, 1 (1998).

\end{thebibliography}
\end{document}